\documentstyle[epsf]{elsart}
\begin{document}

\begin{frontmatter}
  \rightline{HD-THEP-97-35} \rightline{SHEP 97/18} \vskip 1truecm
  
  \title{A lattice determination of QCD field strength correlators}
  \author {Gunnar S.\ Bali}
  \address{\it Department of Physics, The University of Southampton, \\
    Highfield, Southampton SO17 1BJ, England} \author{Nora
    Brambilla$^{~1}$ $^*$ and Antonio Vairo} \footnote{Alexander von
    Humboldt Fellow}
  \address{\it Institut f\"ur Theoretische Physik, Universit\"at Heidelberg\\
    Philosophenweg 16, D-69120 Heidelberg, FRG\\
    $^*$ INFN, Sezione di Milano, Via Celoria 16, 20133 Milano, Italy}
  \maketitle

\begin{abstract}
  \baselineskip=20pt We study field strength correlators in presence
  of a static quark-antiquark pair by use of lattice methods. The
  lattice data have been acquired recently in the context of a
  determination of relativistic corrections to the static interquark
  potential. We extract independent estimates of the form factors of
  two point correlation functions and investigate the effect of higher
  order correlators. Our results confirm the dominance of the two
  point correlators in the regions of intermediate and large quark
  separations, and are compatible with correlation length values
  obtained from different approaches.
\end{abstract}
\end{frontmatter}

\newpage \pagenumbering{arabic}

\section{Introduction}
Understanding the QCD dynamics requires insight into non-trivial QCD
vacuum properties. In their pioneering work, Shifman, Vainshtein and
Zakharov \cite{SVZ} focused on implications of condensation of quarks
and gluons onto elementary particle spectroscopy. Soon, it emerged
that in general not only the condensates had to be considered but also
the space-time dependence of correlation functions as well as a
correlation length \cite{Gromes}. Since then, lattice studies
concentrated on the measurement of the gauge invariant field strength
correlator
\begin{equation}
g^2\langle \phi(0,x) F_{\mu\nu}(x) \phi(x,0) F_{\lambda\rho} (0) \rangle
\,   ,   
\label{uno}
\end{equation}
$$
F_{\mu \nu} = T^a F_{\mu \nu}^a , \quad A_\mu=T^a A_\mu^a , \quad
\phi(x,0) = P \exp \{ ig \int_x^0 dz_\mu A_\mu(z) \} ,
$$
which, at large distances, exhibits an exponential fall-off
behaviour that is controlled by a correlation length $\lambda_g$. The
Stochastic Vacuum Model (SVM) offers a systematic way to treat
non-perturbative QCD effects which enter in terms of gluon field
strength correlation functions \cite{Do88}.  In this approach a
picture emerges in which the correlator (\ref{uno}) plays a basic role
in understanding colour confinement.  It can be decomposed in terms of
two form factors $D$ and $D_1$,
\begin{eqnarray}
&\,& g^2 
\langle \phi(0,x) F_{\mu\nu}(x) \phi(x,0) F_{\lambda\rho} (0) \rangle =
\nonumber\\ 
&\,& \qquad {g^2 \langle F^2(0)\rangle \over 24 N_c (D(0) + D_1(0))} 
\bigg\{ (\delta_{\mu\lambda}\delta_{\nu\rho} - 
\delta_{\mu\rho}\delta_{\nu\lambda})(D(x^2) + D_1(x^2)) 
\nonumber\\
&\,& \quad + (x_\mu x_\lambda \delta_{\nu\rho} - 
x_\mu x_\rho \delta_{\nu\lambda} 
+ x_\nu x_\rho \delta_{\mu\lambda} - x_\nu x_\lambda \delta_{\mu\rho})
{d\over dx^2}D_1(x^2) \bigg\}.
\label{dd1}
\end{eqnarray}
All parameters within the SVM can in principle be derived in a lattice
simulation directly from the QCD Lagrangian.  The simplest linkage
between the analytic model and lattice gauge theory can be found in
the Gaussian approximation to the SVM.  In this approximation
cumulants of order higher than two in the field strength are neglected
in the Wilson loop, such that it is sufficient to evaluate two-point
correlation functions on the lattice.  Within this approximation,
successful predictions ranging from the colour electric field
distribution within the QCD flux tube between two static quarks and
the determination of the QCD static potential to high energy hadron
scattering \cite{svmrev} have been produced.  To our knowledge all
previous lattice evaluations of field strength correlators have
concentrated on determining two point correlators only (see however
\cite{digiasi} for an attempt of incorporating higher order effects)
and relied on Eq. (\ref{dd1}) in the simplest choice of the Schwinger
line factors in connection with the cooling method \cite{Dig97}.

In this paper we have chosen a different approach. We investigate
effects of field strength correlators in presence of a static $q
\bar{q}$ pair to all orders; these in principle contain all physical
information about the flux tube configuration and the $ q \bar{q}$
heavy quark confined system.  With the techniques developed in the
Wilson loop approach (i.e. manifestly gauge-invariant approach to
quark dynamics) the complete gauge-invariant quark-antiquark potential
at order $1/ m^2$ in the quark mass can be obtained from such field
strength correlators \cite{bra94,bra90}. Recently, these analytic
expressions were used to obtain the complete semirelativistic
potential from the lattice \cite{bal97}. Subsequently, the
experimental bottomonium spectrum has been fitted to these QCD
predictions, and a lattice scale as well as the bottom quark mass have
been obtained.

Within the present study, we will use the same (lattice) correlation
functions 1) to investigate the importance of higher order correlation
functions within the range of interquark distances realized ($\sim$
0.1--1 fm) and 2) to extract in this accessible region the behaviour
of the $D$ and $D_1$ functions. Therefore, we do not only aim at a
consistency check of the approximation involved in the definition of
the Gaussian Stochastic Vacuum Model but demonstrate indirectly the
impact of these non-perturbative correlators onto quarkonia spectra.

The paper is organized as follows. In Sec. 2 we introduce the vacuum
expectation values that have been evaluated on the lattice and their
relation to the Wilson loop (and therefore to the potential).
Furthermore, we outline the strategy adopted to identify the
contribution of the higher order correlators; in Sec. 3 we discuss the
numerical results, the validity of the Gaussian approximation in the
SVM and try to extract the large $r$ behaviour of the $D$ and $D_1$
form factors before we present our conclusions in Sec. 4.

\section{Wilson loop and field strength correlation functions}

Given a set of Schwinger lines $\Phi \equiv \{\phi(0,u) = \exp \{ig
\displaystyle \int_{0}^u dz_\mu A_\mu(z)\},\quad u\in E\}$, connecting
points $u$ in Euclidean space $E$ with the origin 0 of the reference
frame, the Wilson loop $\langle W(\Gamma)\rangle\equiv \langle \exp
\{ig \displaystyle \oint_\Gamma dz_\mu A_\mu(z)\} \rangle$ can be
expanded as \cite{Do88,Kampen},
\begin{eqnarray}
{\rm log} \langle W(\Gamma)\rangle &=& 
\sum_{n=0}^\infty {(ig)^n \over n!} 
\int_{S(\Gamma)} dS_{\mu_1\nu_1}(u_1) \cdots  dS_{\mu_n\nu_n}(u_n)
\langle \phi(0,u_1)
\nonumber\\
 &~& \times F_{\mu_1\nu_1}(u_1)\phi(u_1,0) \cdots
\phi(0,u_n) F_{\mu_n\nu_n}(u_n)\phi(u_n,0) \rangle_{\rm cum}.
\label{cumex}
\end{eqnarray}
$S(\Gamma)$ denotes a surface with contour $\Gamma$.\footnote{ Until
  now it has not been completely proven that the series Eq.
  (\ref{cumex}) has a finite convergence radius. We assume that with
  our choice of $\Phi$ a region of convergence exists.} The cumulants
$\langle ~~ \rangle_{\rm cum}$ are defined in terms of path integral
averages over gauge fields,
\begin{eqnarray}
&\,& \langle \phi(0,u_1) F(u_1) \phi(u_1,0) \rangle_{\rm cum} 
\quad = \langle \phi(0,u_1) F(u_1) \phi(u_1,0) \rangle, 
\nonumber\\
&\,& \langle \phi(0,u_1) F(u_1) \phi(u_1,u_2) 
F(u_2) \phi(u_2,0)\rangle_{\rm cum} = 
\nonumber\\
&\,& \quad\quad\qquad\qquad\, \langle \phi(0,u_1) F(u_1) \phi(u_1,u_2) 
F(u_2) \phi(u_2,0)\rangle  
\nonumber\\
&\,& \quad\qquad\qquad - \,\, \langle \phi(0,u_1) F(u_1) \phi(u_1,0)\rangle\,  
\langle \phi(0,u_2) F(u_2)\phi(u_2,0)\rangle,
\nonumber\\
&\,& \qquad\qquad\qquad\qquad\qquad \cdots \quad \quad \cdots \nonumber  
\end{eqnarray}
where $\phi(u_1,u_2)\equiv\phi(u_1,0)\phi(0,u_2)$ is a shorthand
notation.  Notice that since the left-hand side of Eq. (\ref{cumex})
is independent of the set of paths $\Phi$ as well as of the choice of
the surface $S$, all these dependencies are expected to cancel out
after summation of the whole series. However, in general, each
cumulant appearing within the right-hand side will depend on the
choice of $\Phi$.

Let us define
\begin{equation}
{\cal F}_{\mu\nu\lambda\rho}(x;\Gamma) \equiv 
\langle\!\langle g^2F_{\mu\nu}(x)F_{\lambda\rho}(0)\rangle\!\rangle_\Gamma 
- \langle\!\langle gF_{\mu\nu}(x)\rangle\!\rangle_\Gamma 
  \langle\!\langle gF_{\lambda\rho}(0)\rangle\!\rangle_\Gamma.
\label{defF}
\end{equation}
This object is required to compute the ${1/m^2}$ corrections to the
static interquark potential \cite{bra94,bra90}.  The double bracket
$\langle\!\langle ~\cdot~ \rangle\!\rangle_\Gamma$ stands for the
average over gauge fields in presence of the Wilson loop $W(\Gamma)$
$$
\langle\!\langle ~\cdot~\rangle\!\rangle_\Gamma \equiv {\langle
  ~\cdot ~ W(\Gamma) \rangle \over \langle W(\Gamma) \rangle} ;
$$
from now on we will assume that the points $x\equiv({\bf r},t)$ and
$0$ belong to the first and second temporal quark line, respectively
(see Fig. \ref{figw}).

Some particular cases are ($E_i = F_{i4}$, $B_i =
\epsilon_{ijk}F_{jk}/2$):
\begin{eqnarray}
{\cal E}(x;\Gamma) &\equiv&
{\cal F}_{i4i4}(x;\Gamma)= 
\langle\!\langle {\bf E}(x) \cdot {\bf E}(0) \rangle\!\rangle_{\Gamma}  
- \langle\!\langle {\bf E}(x) \rangle\!\rangle_{\Gamma} \cdot 
\langle\!\langle {\bf E}(0) \rangle\!\rangle _{\Gamma},
\nonumber\\
{\cal B}(x;\Gamma) &\equiv&
 {1\over 4}\epsilon_{ilm}\epsilon_{ijk}{\cal F}_{lmjk}(x;\Gamma)=
\langle\!\langle {\bf B}(x) \cdot {\bf B }(0) \rangle\!\rangle_{\Gamma},
\nonumber\\
{\cal C}(x;\Gamma) &\equiv&
{1\over 2}\epsilon_{ijk}{\cal F}_{i4jk}(x;\Gamma)=
\langle\!\langle {\bf E}(x) \cdot {\bf B }(0) \rangle\!\rangle_{\Gamma}  
\nonumber\\
&=&{1\over 2}\epsilon_{ijk}{\cal F}_{jki4}(x;\Gamma) 
= \langle\!\langle {\bf B}(x) \cdot {\bf E }(0) \rangle\!\rangle_{\Gamma},
\nonumber\\
{\cal D}_i(x;\Gamma) &\equiv&
-{1\over 2}\epsilon_{ijk}\epsilon_{klm}{\cal F}_{j4lm}(x;\Gamma)=
-{\bf e}_i\langle\!\langle {\bf E}(x) \wedge {\bf B }(0)
\rangle\!\rangle_{\Gamma}\nonumber\\
&=&-{1\over 2}\epsilon_{ijk}\epsilon_{klm}{\cal F}_{lmj4}(x;\Gamma)=
-{\bf e}_i\langle\!\langle {\bf B}(x) \wedge {\bf E }(0)
\rangle\!\rangle_{\Gamma},
\nonumber
\end{eqnarray}
where the fact that $\langle\!\langle {\bf B}(x) \rangle\!\rangle
_{\Gamma}={\bf 0}$ allows us to disregard some of the disconnected
parts. The equalities in the last two relations are due to PT
invariance. Moreover, due to T invariance, we have ${\cal D}_i({\bf
  r},t;\Gamma)=-{\cal D}_i({\bf r},-t;\Gamma)$\footnote{ For special
  geometries of the contour $\Gamma$ other general properties can be
  derived. In particular, for a static Wilson loop we obtain
  $\langle\!\langle E_i(x)\rangle\!\rangle_{\Gamma}=
  \displaystyle{r_i\over r}\displaystyle{\partial V_0(r) \over
    \partial r}= -\langle\!\langle E_i(0)\rangle\!\rangle_{\Gamma}$
  where $V_0(r)$ denotes the static potential between colour sources,
  separated by a spatial distance $r$. On the lattice this symmetry
  holds only for large $r$, such that we prefer to determine
  $\langle\!\langle E_i\rangle\!\rangle_{\Gamma}$ by use of the
  relation $\langle\!\langle {\bf E}(x) \rangle\!\rangle_{\Gamma}
  \cdot \langle\!\langle {\bf E}(0) \rangle\!\rangle _{\Gamma}=
  {\displaystyle \lim_{t\rightarrow\infty}} \langle\!\langle {\bf
    E}({\bf r},t)\cdot {\bf E}(0) \rangle\!\rangle _{\Gamma}$.}.

The function $\cal F$ can be exactly expressed as a functional
derivative of the Wilson loop average \cite{bra94,bra97}:
\begin{equation} 
{\cal F}_{\mu\nu\lambda\rho}(x;\Gamma) = 
{\delta \over \delta S_{\mu\nu}(x)} 
{\delta \over \delta S_{\lambda\rho}(0)}
{\rm \log} \langle W(\Gamma) \rangle.
\label{der2}
\end{equation}

In Refs. \cite{bra94,bra97} the whole semirelativistic $ q\bar{q}$
potential was obtained as a gauge-invariant function of the first and
second functional derivative on the logarithm of the Wilson loop.
These derivatives correspond to the vacuum expectation values of one
or two field strengths insertions in the static Wilson loop
\cite{bra90} that have been recently evaluated on the lattice
\cite{bal97}.

Taking Eq. (\ref{cumex}) into account, we obtain,
\begin{equation}
{\cal F}_{\mu\nu\lambda\rho}(x;\Gamma) = 
g^2\langle \phi(0,x) F_{\mu\nu}(x) \phi(x,0) F_{\lambda\rho} (0) \rangle  
+ {\cal R}_{\mu\nu\lambda\rho}(x;\Gamma),
\label{FF}
\end{equation} 
where $\cal R$ contains all contributions from cumulants of order
higher than two:
\begin{eqnarray}
{\cal R}_{\mu\nu\lambda\rho}(x;\Gamma) &=& 
\sum_{n=3}^\infty 
{(ig)^n\over n!} \sum_{\{ {\rm all \, perm.} (i,j) \} } 
\int_{S(\Gamma)} \left[ \prod_{k \neq i,j}^n dS_{\mu_k\nu_k}(u_k)\right] 
\nonumber\\
&\times&  
\langle \phi(0,u_1) F_{\mu_1\nu_1}(u_1) \cdots 
F_{\mu\nu}(u_i = x) \cdots  
\nonumber\\
&\times& 
\quad\quad \cdots F_{\lambda\rho}(u_j=0) \cdots  
F_{\mu_n\nu_n}(u_n)\phi(u_n,0) \rangle_{\rm cum}.
\label{rest}
\end{eqnarray} 

In the so-called Stochastic Vacuum Model (SVM) it is assumed that for
large distances the bilocal cumulant $\langle \phi(0,x) F_{\mu\nu}(x)
\phi(x,0) F_{\lambda\rho} (0) \rangle$ is the dominant contribution to
${\cal F}$ in Eq. ({\ref{FF}) \cite{svmrev,bra97,Si88}. Higher order
  cumulants are expected to be small corrections to the relevant
  non-perturbative parameters (like the string tension) that can be
  derived from the bilocal one. We have already pointed out that this
  assumption has been successfully tested in potential theory and soft
  high energy scattering (for some reviews see Ref. \cite{svmrev}).
  However, in the present work we aim at a first principles check of
  this assumption on the lattice. The adopted strategy is the
  following.  Lorentz invariance is only violated by the contour
  $\Gamma$ within Eq. (\ref{FF}). Therefore, deviations of $\cal F$
  from Lorentz invariance have to be attributed to the higher order
  cumulants $\cal R$ on the right hand side. We interpret the Lorentz
  invariant part of the data as $\langle \phi(0,x) F_{\mu\nu}(x)
  \phi(x,0) F_{\lambda\rho} (0) \rangle$ and parameterize it in terms
  of two form factors. These can subsequently be compared with other
  lattice determinations of these quantities which have been obtained
  by use of the so-called cooling method \cite{Dig97}.  We interpret
  deviations between our data points and continuous parameterizations
  as an estimate for the size of higher order corrections.

\begin{figure}[htb]
  \vskip 0.8truecm \makebox[2.5truecm]{\phantom b} \epsfxsize=10truecm
  \epsffile{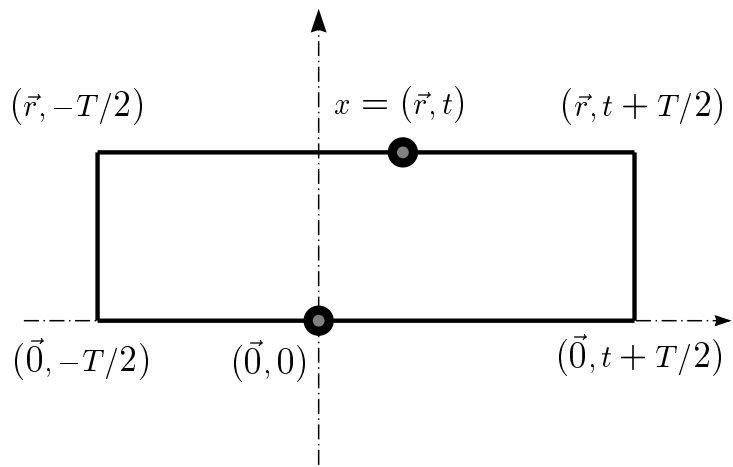}
\caption{{{\it The rectangular Wilson loop $\Gamma$ with two field 
      strength insertions.}}}
\label{figw}
\vskip 0.8truecm
\end{figure}

Let us focus on the bilocal cumulant $\langle \phi(0,x) F_{\mu\nu}(x)
\phi(x,0) F_{\lambda\rho} (0) \rangle$ given in Eq. (\ref{dd1}).  For
our purposes it is more convenient to define the functions:
\begin{eqnarray}
D_\perp(x^2) &\equiv&  
{g^2\langle F^2(0)\rangle \over 24 N_c (D(0) + D_1(0))} (D(x^2) + D_1(x^2)), 
\label{dperp}\\ 
D_*(x^2) &\equiv& {g^2\langle F^2(0)\rangle \over 24 N_c (D(0) + D_1(0))} 
x^2 {d\over dx^2}D_1(x^2).
\label{dstar}
\end{eqnarray}
Note that from Eq. (\ref{dperp}) the gluon condensate is given by
$G_2\equiv \langle g^2 F^2(0)/4 \pi^2\rangle = D_{\perp}(0) \,6 N_c /
\pi^2$.  The cumulant expansion Eq.~(\ref{cumex}) takes into account
all perturbative as well as all non-perturbative contributions to the
Wilson loop.  In particular, for what concerns the bilocal cumulant of
Eq. (\ref{dd1}), the origin of $D$ is purely non-perturbative (with a
typical behaviour $\sim {\displaystyle e^{-|x|/\lambda_g}}$ for large
$x$) while $D_1$ contains perturbative ($\sim 1/x^4$) as well as
non-perturbative contributions.

Some special cases of Eq. (\ref{dd1}) are,
\begin{eqnarray}
g^2\langle \phi(0,x) F_{i4}(x) \phi(x,0) F_{i4} (0) \rangle &=& 
3 (D_\perp(x^2) + D_*(x^2)) 
\nonumber\\
&-& 2 {{\bf r}^2 \over x^2} D_*(x^2), 
\label{EE}\\
\epsilon_{ilm}\epsilon_{ijk}
g^2\langle \phi(0,x) F_{lm}(x) \phi(x,0) F_{jk} (0) \rangle &=& 
12 D_\perp(x^2) + 8 { {\bf r}^2 \over x^2} D_*(x^2), 
\label{BB}\\
\epsilon_{ijk}
g^2\langle \phi(0,x) F_{i4}(x) \phi(x,0) F_{jk} (0) \rangle &=& 0,
\label{EB}\\
\epsilon_{ijk}
g^2\langle \phi(0,x) F_{jk}(x) \phi(x,0) F_{i4} (0) \rangle &=& 0,
\label{BE}\\
\epsilon_{ijk}\epsilon_{klm}
g^2\langle \phi(0,x) F_{j4}(x) \phi(x,0) F_{lm} (0) \rangle &=&
-4{ r_i t \over x^2} D_*(x^2),
\label{EB2}\\
\epsilon_{ijk}\epsilon_{klm}
g^2\langle \phi(0,x) F_{lm}(x) \phi(x,0) F_{j4} (0) \rangle &=&
-4{ r_i t \over x^2} D_*(x^2).
\label{BE2}
\end{eqnarray}

With the above definitions, we can rewrite Eq. (\ref{FF}) as
\begin{eqnarray}
{\cal F}_{\mu\nu\lambda\rho}(x;\Gamma) &=&
(\delta_{\mu\lambda}\delta_{\nu\rho} - 
\delta_{\mu\rho}\delta_{\nu\lambda})D_\perp(x^2)  
\nonumber\\
&+& {1\over x^2} (x_\mu x_\lambda \delta_{\nu\rho} 
- x_\mu x_\rho \delta_{\nu\lambda} + x_\nu x_\rho \delta_{\mu\lambda} 
- x_\nu x_\lambda \delta_{\mu\rho}) D_*(x^2)\label{RR}\\\nonumber
&+& {\cal R}_{\mu\nu\lambda\rho}(x;\Gamma),
\label{dd11}
\end{eqnarray}
and, in particular,
\begin{eqnarray}
{\cal E}(x;\Gamma)  &=&
 3 ( D_\perp(x^2) + D_*(x^2)) - 2 {r^2 \over x^2} D_*(x^2)+
{\cal R}_{i4i4}(x;\Gamma),
\label{EERb}\\
{\cal B}(x;\Gamma) &=&
3 D_\perp(x^2) + 2 { r^2 \over x^2} D_*(x^2)
+{1\over 4}\epsilon_{ilm}\epsilon_{ijk}{\cal R}_{lmjk}(x;\Gamma),
\label{BBRb}\\
{\cal C}(x;\Gamma)&=&
\frac{1}{2}\epsilon_{ijk}{\cal R}_{i4jk}(x;\Gamma),
\label{EBRb}\\
{\cal D}(x, \Gamma) &\equiv& 
{r_i\over r}{\cal D}_i(x;\Gamma)={2rt\over x^2}D_*(x^2)
-{1\over 2}{r_i\over r}\epsilon_{ijk}\epsilon_{klm} {\cal R}_{j4lm}(x;\Gamma).
\label{EB2Rb} 
\end{eqnarray}

The lattice data have been produced in order to evaluate the complete
$1/m^2$ corrections to the static quark-antiquark potential
\cite{bal97}. The loop $\Gamma$ is the rectangle depicted in Fig.
\ref{figw} in the limit of large separation in Euclidean time between
the coordinates of the field strength insertions (ears) and the
spatial closures of the Wilson loop ($T \gg 0,t$).  In practice, this
limit has been realized by ``smearing'' the spatial connections within
the Wilson loop, i.e.\ by substituting the straight paths by a
suitable linear combination of paths that maximizes the overlap with
the $q\overline{q}$ ground state (For details see Refs.
\cite{bal97,bal96}). The data have been non-perturbatively
renormalized by use of the Huntley--Michael method \cite{hun87}.

The lattice results provide information on the functions ${\cal E}$,
${\cal B}$ and ${\cal D}$. However, not all possible choices of $\cal
F$ have been realized, such that no data on $\cal C$ is available.  In
the next section we will determine the functions $D_\perp$ and $D_*$,
assuming that $\cal R$ in Eqs.~(\ref{EERb}), (\ref{BBRb}) and
(\ref{EB2Rb}) can be neglected in the Lorentz-symmetric region. Since
we have more relations than functions we can also provide an
independent consistency check of this assumption. In doing so we
obtain two results.  Firstly, as long as the data supply us with
space-time symmetric functions $D_\perp$ and $D_*$ we can consistently
assume that contributions from higher order cumulants can be
neglected. In fact ${\cal E}$, ${\cal B}$ and $\cal D$, depending on
$\Gamma$, are in general not space-time isotropic functions, whereas
$D_\perp$ and $D_*$ are.  On the other hand, the only source of
anisotropy is $\cal R$.  Secondly, in the region where the bilocal
cumulants are established to be the dominant contributions, we obtain
a parameterization of the form factors that can be compared to results
already existing in the literature\footnote{The definition of the
  bilocal cumulant is not unique but depends on the set of connecting
  paths $\Phi$. However, we expect the large distance behaviour of the
  cumulants to be universal, in particular the correlation length.}.

\section{Results}

{}From Eqs. (\ref{EERb}), (\ref{BBRb}) and (\ref{EB2Rb}) we have:
\begin{eqnarray}
2 r t D_*(x^2) &=& x^2 {\cal D}(x;\Gamma) +\cdots,   
\label{l4}\\
D_{\perp}(x^2) + {1\over 2} D_*(x^2) &=&
{1\over 6} \left({\cal B}(x;\Gamma)+{\cal E}(x;\Gamma)\right) + \cdots,
\label{l2}\\
(x^2 -4 t^2) D_*(x^2) &=& x^2 \left({\cal B}(x;\Gamma)
-{\cal E}(x;\Gamma)\right)
+\cdots
\label{l1}.
\end{eqnarray}
By inverting these relations we obtain $D_*$ and $D_\perp$ for various values 
of ${\bf r}$ and $t$. In order to check space-time symmetry 
we will consider several combinations of  ${\bf r}$ and $t$ which yield
similar $x^2 = r^2 + t^2$. The required correlation functions 
$\cal E$, $\cal B$ and $\cal D$ have been measured on the lattice 
in the context of the extraction of relativistic corrections to the 
static interquark potential \cite{bal97}. The lattice simulations 
(with Wilson action, within the valence quark approximation to QCD) 
have been performed at the inverse lattice couplings $\beta=6.0$ and 
$\beta=6.2$ on $16^4$ and $32^4$ lattices, respectively. From a fit to 
the experimental bottomonium spectrum, we obtain inverse lattice spacings 
of $a^{-1}=2.14$ GeV and $a^{-1}=2.94$ GeV ($a=0.092$ fm and $a=0.067$ fm), 
respectively. In what follows, we will neglect systematic errors on these 
scale estimates from sea quark effects and higher order relativistic 
corrections \cite{bal97}.

\begin{figure}[htb]
\vskip 0.8truecm
\makebox[0.1truecm]{\phantom b}
\epsfxsize=13truecm
\epsffile{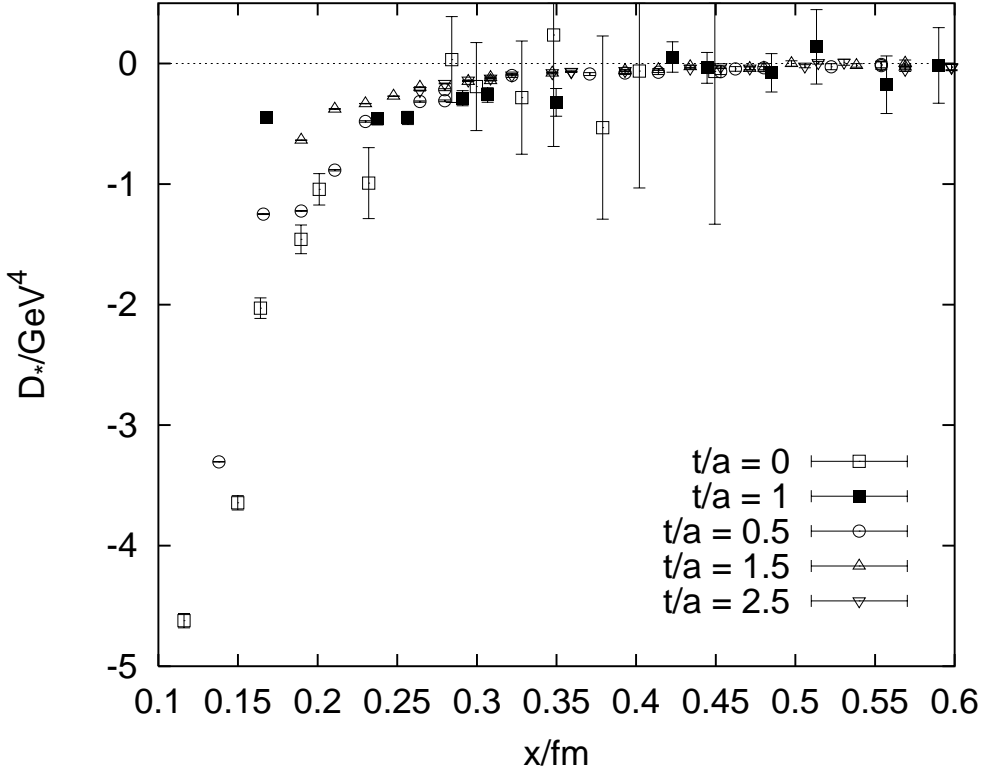}
\vskip 0.3truecm
\caption{{{\it Estimates of $D_*$ from Eqs. (\ref{l1}) (squares) and
(\ref{l4}) (other symbols) at $\beta = 6.0$.}}}
\label{figdsf}
\vskip 0.8truecm
\end{figure}

\begin{figure}[htb]
\vskip 0.8truecm
\makebox[0.1truecm]{\phantom b}
\epsfxsize=13truecm
\epsffile{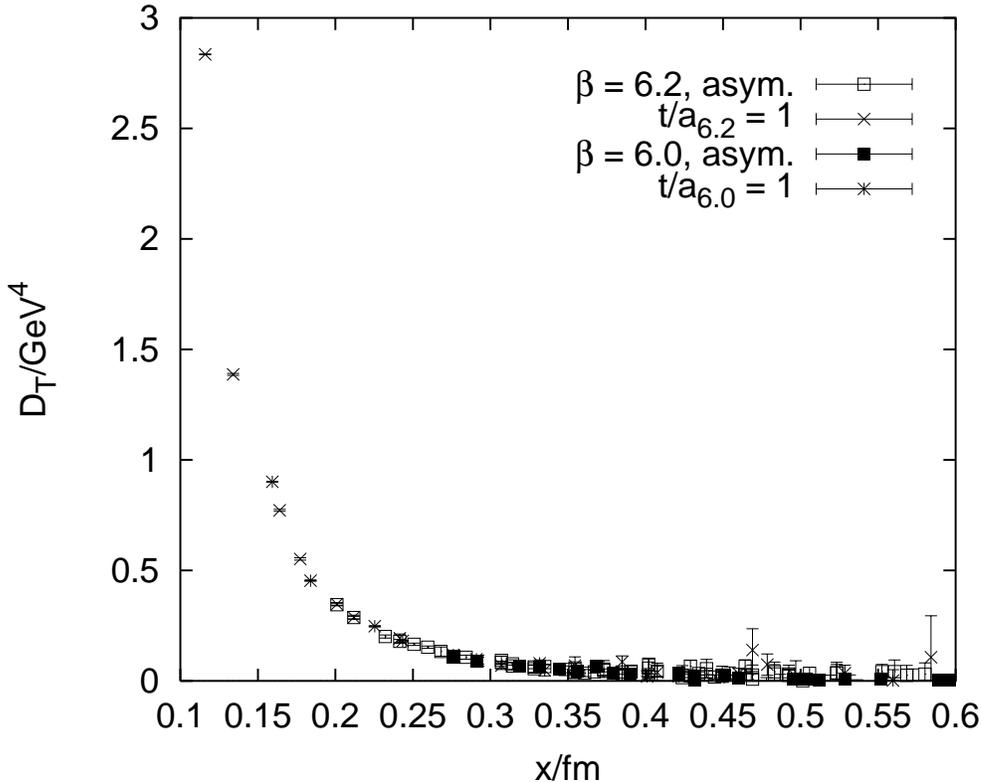}
\vskip 0.3truecm
\caption{{{\it Estimates of
$D_{\perp}$ from Eqs. (\ref{l2}) and (\ref{l4})
at $\beta = 6.0$ and $\beta=6.2$
in comparison to the asymptotic values (squares).}}}
\label{figdpf}
\vskip 0.8truecm
\end{figure}

Estimates of $D_*$ and $D_\perp$ from Eqs.~(\ref{l4})--(\ref{l1})
are displayed in Figs.~\ref{figdsf} and \ref{figdpf}, respectively. 
The data points at integer values of $t/a$ in Fig.~\ref{figdsf}
have been obtained from Eq.~(\ref{l1}) while
the data at half integer values are obtained from Eq.~(\ref{l4}).
We observe that for large $x$ all data collapse onto a universal curve.
The data from Eq. (\ref{l1}) (squares) provides an 
independent consistency check with respect to Eq.~(\ref{l4}).  
For small $x$, significant deviations from a universal curve are visible
in both cases. These deviations signal that higher order cumulants become 
important which is simply an algebraic consequence 
of Eqs.~(\ref{l4}) and (\ref{l1}): the function multiplying $D_*$ 
in Eq.~(\ref{l4}) vanishes for small values of $r = \sqrt{x^2-t^2}$
and the function multiplying $D_*$ in Eq.~(\ref{l1}) vanishes at 
the points $x = 2t$. For the values of $t$ taken into 
account, both functions are nearly zero for small $x$. Therefore,
in this region, contributions of bilocal cumulants
to $\cal D$ and ${\cal E} - {\cal B}$ are suppressed with respect to
those from higher order cumulants which, as discussed above, are in general 
not space-time symmetric. We find that the observed behaviour is qualitatively 
reproduced at $\beta=6.2$. In fact, data sets obtained at pairs of
similar $t$-values (in physical units) exhibit approximate scaling, such that
the data points of the figure seem to be relevant to continuum
physics, rather than being mere lattice artifacts.

$D_{\perp}$ has been obtained by substituting the result on $D_*$ from
Eq. (\ref{l4}) (which provides us with more precise data  
than Eq. (\ref{l1})) into Eq. (\ref{l2}).
Contrary to the estimates on $D_*$, in this case 
deviations of data points from an asymptotic curve are small, even at small $x$. 
In Fig.~\ref{figdpf} we display a comparison between the $t=a$ data, obtained 
at $\beta=6.0$ and $\beta=6.2$. The results do not only weakly depend on $t$ but are 
almost independent of the lattice spacing $a$, even at these small $t$ values.

Within the present statistical errors, universality of the curves is obtained from
$r_{\min}\geq\sqrt{5}a$ and $t_{\min}\geq 3.5a$ ($t_{\min}\geq 2.5a$ 
at $\beta=6.0$) for $D_*(x)$ and $r_{\min}\geq\sqrt{5}a$ and $t_{\min}\geq 2a$ 
for $D_{\perp}(x)$ onwards. In physical units this distance corresponds
to $x_{\min}\approx 0.3$ fm for $D_*$ and $x_{\min}\approx 0.2$ fm 
for $D_{\perp}$, respectively. 

\begin{figure}[htb]
\vskip 0.8truecm
\makebox[0.1truecm]{\phantom b}
\epsfxsize=13truecm
\epsffile{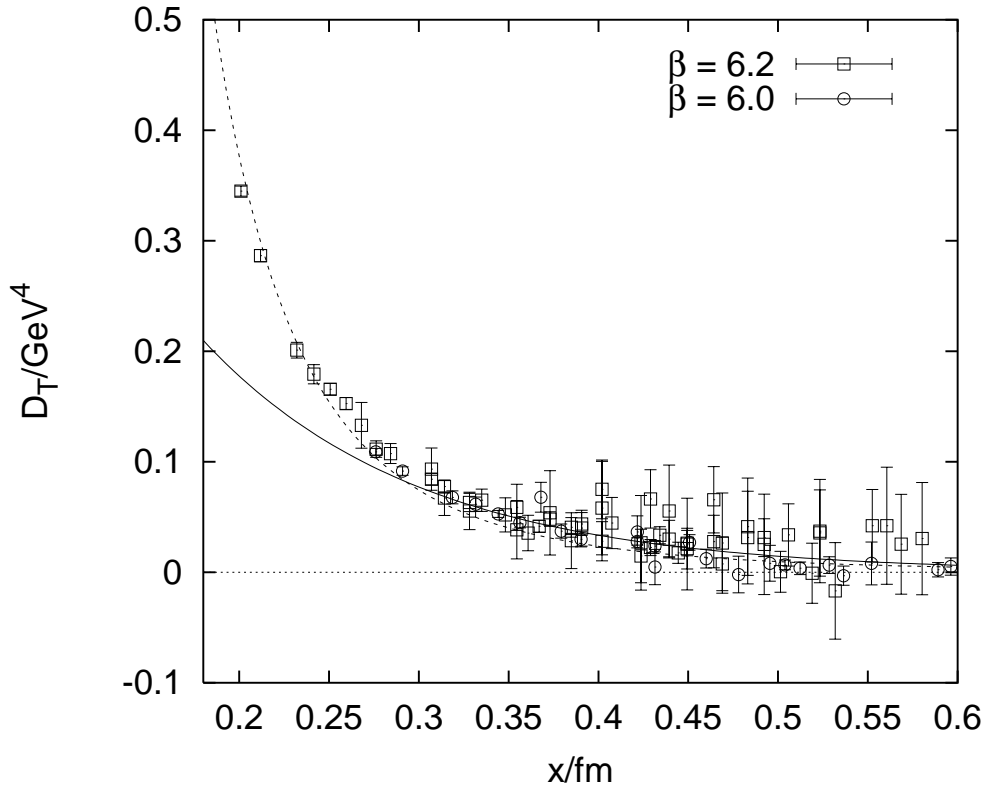}
\vskip 0.3truecm
\caption{{{\it $D_{\perp}$ with a one exponential fit to the $\beta=6.0$ 
data (solid curve) and the curve $0.4/x^4$ (dashed).}}}
\label{figdp}
\vskip 0.8truecm
\end{figure}

\begin{figure}[htb]
\vskip 0.8truecm
\makebox[0.1truecm]{\phantom b}
\epsfxsize=13truecm
\epsffile{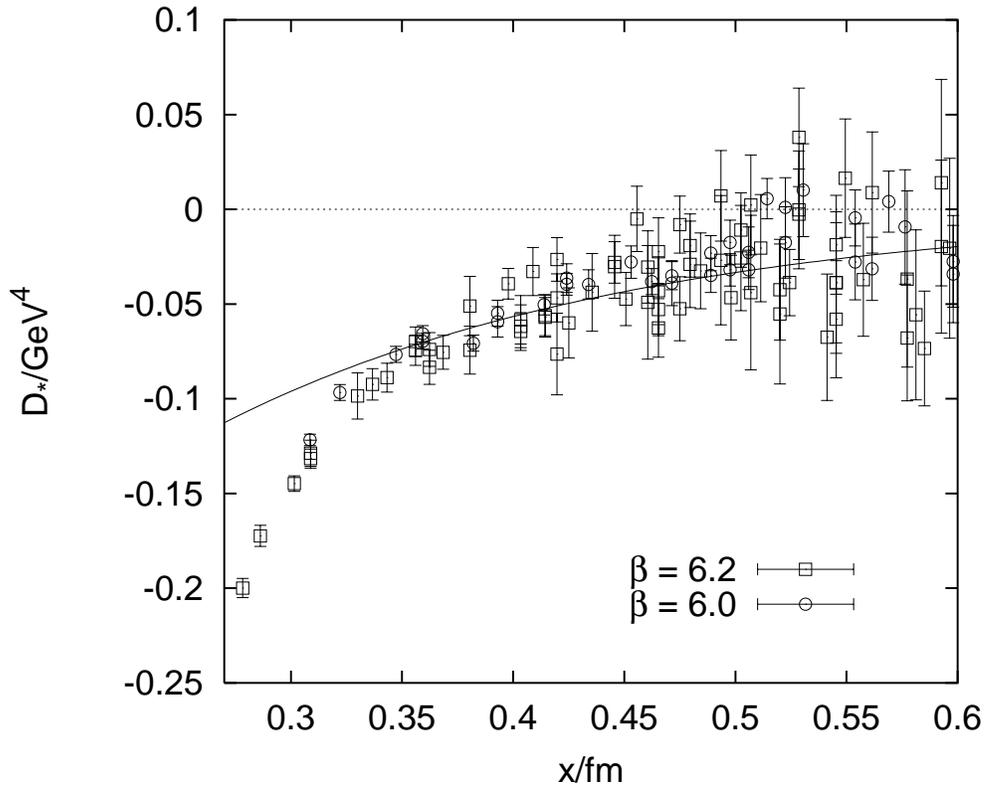}
\vskip 0.3truecm
\caption{{{\it $D_*$ with a one exponential fit to the $\beta=6.0$ data.}}}
\label{figds}
\vskip 0.8truecm
\end{figure}

In Fig. \ref{figdp} all asymptotic data on $D_{\perp}$ is displayed,
together with a
one-ex\-po\-nen\-ti\-al fit to the $\beta=6.0$ data of the form (solid line)
\begin{equation}
D_{\perp}(x^2)=A \exp(-|x|/\lambda_a),
\end{equation}
The fit parameters remain stable for fit ranges $x_{\min} > 0.3$ fm 
(see Table \ref{et2}). Although even for smaller $x_{\min}$, 
exponential fits yield reasonable $\chi^2$ values, the fit parameters turn 
out to be unstable under variation of the fit range. Within errors 
the parameters obtained at the two lattice spacings agree with each other. 
We quote as a final result the more precise value
obtained at $\beta=6.0$,
\begin{equation}
\lambda_a=0.120^{+0.009}_{-0.012} \,\mbox{fm}.
\end{equation}
In addition, a curve $c/x^4$, which is the functional form one would expect 
from perturbation theory, with $c = 0.4$ is plotted  in order to show 
the dominance of the perturbative regime in the short range region. 

\begin{table}
\caption{Fits to $D_{\perp}$.}
\label{et2}
\begin{tabular}{ccccccc}
$\beta$&$x_{\min}/a$&$\lambda_a/$fm&$A/$GeV$^4$&$\chi^2/N_{DF}$\\\hline
6.0&3.46&$0.120^{+09}_{-12}$&$0.94^{+32}_{-16}$& 0.68\\
6.2&4.69&$0.153^{+22}_{-35}$&$0.56^{+47}_{-15}$& 0.80\\\hline
\end{tabular}
\end{table}

In Fig. \ref{figds} the asymptotic data on $D_*$ is plotted, together with
an exponential fit  of the form
\begin{equation}
D_*(x^2)= -B \exp(-|x|/\lambda_b),
\end{equation}
to the $\beta=6.0$ data. The fit parameters remain stable for fit ranges
$x_{\min}>0.35$ fm (see Table \ref{et1}). Again, within errors the parameters 
obtained at the two lattice agree with each other. We end with the
$\beta=6.0$ estimate,
\begin{equation}
\lambda_b=0.189^{+0.013}_{-0.029} \,\mbox{fm}.
\end{equation}
Note that $\lambda_a$ and $\lambda_b$ are lower limits on the
corresponding asymptotic correlation length(s) while $A$ and $B$
are upper limits on the asymptotic amplitudes.

\begin{table}
\caption{Fits to $D_{*}$.}
\label{et1}
\begin{tabular}{ccccccc}
$\beta$&$x_{\min}/a$&$\lambda_b/$fm&$B/$GeV$^4$&$\chi^2/N_{DF}$\\\hline
6.0&3.77&$0.189^{+13}_{-29}$&$0.47^{+20}_{-06}$& 1.25\\
6.2&4.92&$0.249^{+16}_{-54}$&$0.33^{+17}_{-04}$& 1.47\\\hline
\end{tabular}
\end{table}

\section{Conclusions}

In this letter we have shown that the lattice data for the long range 
behaviour of the two-field strength insertion on a rectangular Wilson loop 
are compatible with a two point cumulant approximation. 
This approximation is crucial in the so-called Stochastic Vacuum Model 
of QCD. Moreover our data are compatible with an exponential 
fit $\sim e^{-|x|/\lambda_g}$ with a gluonic correlation length $\lambda_g$ 
of around $0.1$--$0.2$ fm.
These findings are in agreement with the quenched results 
obtained in \cite{Dig97} by means of lattice simulations with the cooling 
method. In general one might expect the cooling method, that removes short 
range fluctuations, to have the tendency of underestimating masses and 
thus to overestimate correlation lengths. On the other hand our 
results are more sensitive to the perturbative regime which certainly
dominates the region in which
the exponential curves of Fig. \ref{figdp} and Fig. \ref{figds} 
deviate from the data (typically for distances shorter than 0.3 fm). 
For the moment being, our data suffer under much bigger statistical errors 
than those obtained after cooling. This is natural since we do not cut off 
short range fluctuations. However, we obtain first independent estimates of
the form factors. Unlike cooling, the Huntley--Michael lattice renormalization 
does not affect transfer matrix elements. Moreover, it has been checked 
\cite{bal97} against the exact continuum Gromes \cite{gro84} and 
Barchielli--Brambilla--Prosperi \cite{bra90} relations. Another advantage of 
the present method is that the two point cumulant has been extracted from
quantities that have a direct physical interpretation and play a role in 
heavy quarkonia spectroscopy. We have shown that an evaluation 
of the bilocal cumulant from existing lattice data is possible with the 
proposed method. We expect that in the near future more selective fits can 
be performed on more precise data.

\vskip 1truecm
{\bf Acknowledgments} 
G. B. has been supported by DFG grant No.\ Ba 1564/3-1 and PPARC grant 
No.\ GR/K55738. He expresses thanks to his collaborators Armin Wachter 
and Klaus Schilling. N. B. is pleased to acknowledge useful discussions 
with A. Di Giacomo. Computations have been performed on the Connection 
Machines CM-5 of the Institut f\"ur Angewandte Informatik in Wuppertal and 
the GMD in St.\ Augustin. We appreciate support from the EU  (grant Nos.
\ SC1*-CT91-0642 and CHRX-CT92-00551) and the Deutsche Forschungsgemeinschaft 
(DFG grant Nos.\ Schi 257/1-4 and Schi 257/3-2).

\end{document}